# The Role of Engagement, Honing, and Mindfulness in Creativity

Liane Gabora and Mike Unrau
*University of British Columbia, Canada*

**Abstract**  As both our external world and inner worlds become more complex, we are faced with more novel challenges, hardships, and duress. Creative thinking is needed to provide fresh perspectives and solve new problems. Because creativity can be conducive to accessing and reliving traumatic memories, emotional scars may be exacerbated by creative practices before these are transformed and released. Therefore, in preparing our youth to thrive in an increasingly unpredictable world, it could be helpful to cultivate in them an understanding of the creative process and its relationship to hardship, as well as tools and techniques for fostering not just creativity but self-awareness and mindfulness. This chapter is a review of theories of creativity through the lens of their capacity to account for the relationship between creativity and hardship, as well as the therapeutic effects of creativity. We also review theories and research on aspects of mindfulness attending to potential therapeutic effects of creativity. Drawing upon the creativity and mindfulness literatures, we sketch out what an introductory 'creativity and mindfulness' module might look like as part of an educational curriculum designed to address the unique challenges of the 21st Century.

Please direct correspondence regarding the manuscript to
Liane Gabora
Email: liane [dot] gabora [at] ubc [dot] ca
Dept. of Psychology, University of British Columbia, Fipke Centre for Innovative Research, 3247 University Way, Kelowna BC V1V 1V7, CANADA

**Liane Gabora**, PhD, is a Professor in the Psychology Department at the Okanagan Campus of the University of British Columbia, Canada. Her research focuses on the mechanisms underlying creativity, and how creative ideas—and culture more generally—evolve, using both computational modeling and empirical studies with human participants. She has almost 200 articles published in scholarly books, journals, and conference proceedings, has procured over one million dollars in research funding, supervised numerous graduate and undergraduate students, and given talks worldwide on creativity and related topics. Her research on creativity is informed by her own experiences creating literature. She has a short story published in *Fiction*, another forthcoming in *Fiddlehead,* and has a novel underway titled *Quilandria* that merges her scholarly and creative writing interests. Her paintings have been exhibited in the United States at *Anastasia's Asylum Coffee House* (Los Angeles), *Ten Women Gallery* (Los Angeles), and *Java Joe's Coffee House* (Santa Fe). Her animated short film titled "*Self-referential Face*" was shown at the *Artificial Life VI Conference*, University of California, Los Angeles. Her electronic music composition *Stream Not Gone Dry* was performed at Royce Hall, University of California, Los Angeles (a piano version can be found at https://people.ok.ubc.ca/lgabora/artistic_files/Gabora-Liane-Stream.wav).

**Mike Unrau, MFA,** is a PhD student in Interdisciplinary Graduate Studies at the University of British Columbia (Okanagan Campus), Canada. He is studying creativity and social innovation, focusing on how creative mindfulness impacts collective trauma towards societal change. He is currently adjunct faculty with the University of Calgary and Mount Royal University, working in field education and simulated educational experiences, as well as social-based theatre and creativity. He has held international fellowships, given lectures and conference presentations, conducted workshops and led research projects in different parts of the world, including a pre-social lab in India. These days Mike's passion is *Somativity,* a physical movement mindfulness approach he developed after living in a Buddhist monastery (Thailand and Canada) and co-founding a physical theatre company (Calgary, AB). He has published findings on somatic awareness as well as creativity, is a published poet, and a competition finalist in song writing and internationally in photography. He is trained in expressive arts, a mindfulness approach called *Living Inquiries*, and is a certified *Transformational Arts* facilitator.

**1.1 Introduction**
With intelligence increasing across generations (a phenomenon referred to as the Flynn Effect), our networks of thoughts and ideas are taking on more diverse, complex structures. As one effect, our minds are traveling down less-trodden, newer paths. Some thought-paths lead to revolutionary innovations, heart-wrenching tunes, and riveting movies. There are other thought-paths that lead to ruminations about the past, and fears about the future, all of which may play a role in anxiety or depression. Others lead to ever-subtler ways of manipulating each other, probing and bringing to light repressed and offensive parts of ourselves.

Thus, as both our external world and inner worlds become more complex, we are faced with novel challenges, hardships, and duress. Creative thinking is needed to provide fresh perspectives and solve new problems, but since creativity can be conducive to accessing and reliving traumatic memories, emotional scars may be exacerbated by creative practices before they are transformed and released. This suggests that in order to prepare our youth to thrive in an increasingly unpredictable world, it could be helpful to cultivate in them an understanding of the creative process and its relationship to hardship, as well as tools and approaches for fostering, not just creativity, but self-awareness and mindfulness. As a working definition, *mindfulness* is the awareness of what is happening presently, by paying attention to our experience and the novel distinctions of it we actively draw upon, without judgement (Kabat-Zinn, 2003; Haller, Bosma, Kapur, Zafonte, Langer, 2016). Mindfulness is essentially creative, in that as we experience life mindfully, what we notice is new to us as a fresh perspective (Langer, Moldoveanu, 2000).

This chapter begins with a brief overview of a few theories of creativity. Our focus is on how, and to what extent, they address the relationship between creativity and hardship, as well as the well-documented therapeutic impact of creative engagement. Next, we investigate some theoretical aspects of mindfulness, again attending to its relationship to hardship and well-being. Finally, we sketch out the basics of what a 'creativity and mindfulness' module might look like.

**1.2 Hardship and the Therapeutic Effects of Creativity**
It is widely believed that creativity is fostered by a warm, supportive, nurturing, and trustworthy environment conducive to self-actualization (Maslow, 1971; Rogers, 1959). However, there is a negative correlation between creativity and parental warmth (Siegelman, 1973) and a high incidence of early parental loss in eminent creators (Eisenstadt, 1978). More generally, childhood adversity is believed to be a developmental antecedent of eminent creativity (MacKinnon, 1962; Rhue & Lynn, 1987; Simonton, 1994), perhaps in part because of a relationship between adversity and diversifying experiences (Damian & Simonton, 2015). The relationship between creativity and hardship is not restricted to childhood adversity; for example, stories written by adults in response to mildly threatening stimuli were rated as more creative than stories responding to non-threatening stimuli (Riley & Gabora, 2012). Thus, a

Nevertheless, creativity is believed to have therapeutic effects (Barron, 1963; Forgeard, 2013). It can be intrinsically rewarding (Gruber, 1995; Kounios & Beeman, 2014; Martindale, 1984). The creative process can at times be frustrating and draining and involve working through negative material. However, there is evidence that high levels of creativity are correlated with positive affect (Hennessey & Amabile, 2010) and the ability to manage intense feelings (Moon, 1999). Clinical practitioners of art therapy note that imagery and creative engagement can deepen communication between client and therapist (Moon, 2009). Art therapy can also enhance self-understanding, in addition to facilitating the process of finding healthier ways of handling situations and interacting with others (Dunn-Snow & Joy-Smellie, 2000; Riley, 1999). By providing access to issues that are difficult to verbalize, art therapy can bring these to the surface in a nonverbal form or provide a springboard for discussion (Malchiodi, 2007). There is also evidence that creativity can enhance one's sense of self (Garailordobil & Berrueco, 2007; MacKinnon, 1962).

Therapeutic effects of creativity may also stem from the capacity of therapy to enhance feelings of connection to, and appreciation by, others. In the verification of a creative work, the creator generates an internal context for the idea that encompasses a typical individual who will encounter the work. For an inventor this (what?) might involve developing a working prototype. For an artist it might involve arranging artworks for showing at a gallery. By finding a form for the idea that is palatable (e.g., comprehensible or intriguing) to others, one's worldview merges with and expands those of others. To the extent that a creative product responds to universal features of worldviews, it may have a healing effect on others. Creative products are felt to be a highly personal form of self-disclosure; self-disclosure has therapeutic value (Pennebaker, 1997) and even beneficial effects on the immune system (Pennebaker, Kiecolt-Glaser, & Glaser, 1998). Since creative people often feel disconnected from others because they defy convention (Sternberg & Lubart, 1995; Sulloway, 1996), the benefits of creative self-disclosure may be mediated by an enhanced sense of belonging.

**1.3 Theories of Creativity**
To what extent do theories of creativity incorporate and account for (1) the relationship between creativity and hardship, and (2) the transformative and sometimes therapeutic effects of creativity? In this section we address these questions.

A starting point for much research into creativity is Wallas' (1926) classification of the creative process into a series of stages. The first stage is preparation, which involves obtaining background knowledge relevant to the problem, and its history, such as any past attempts, or preconceptions regarding how to solve it. It also involves conscious, focused work on the problem. The second stage is incubation—unconscious processing of the problem that continues while one is engaged in other tasks. The first and second stages may be interleaved. Wallas proposed that after sufficient preparation and incubation, the creative process is often marked by a sudden moment of illumination or insight during which the creator glimpses a way of going about the task, which may require substantial work to bring to completion. The final phase is referred to as verification. This involves not just fine-tuning the work and making certain it is works not just in theory but in in practice, but putting it in a form that can be understood and appreciated by others.

While early research supported Wallas' classic four-stage theory of creativity, subsequent studies, in particular those examining the need for incubation, did not support Wallas' theory (Eindhoven & Vinacke, 1952). One criticism of Wallas' theory is that it is merely descriptive and thus fails to explain how or why the stages occur. More importantly with respect to our purposes, it does not address the relationship between creativity and hardship or the therapeutic impact of creativity. However, variants of Wallas' theory have continued to serve as a platform for theoretical and empirical research on creativity.

**1.4 Heuristic Search**

Inspired by the metaphor of the mind as a computer (or computer program), early research on creativity focused on the notion of heuristic search. In heuristic search, [rules of thumb guide the inspection of different states within a particular state space (i.e., a set of possible solutions) until a satisfactory solution is found (Eysenck, 1993; Newell, Shaw, & Simon 1957; Newell & Simon, 1972). In heuristic search, the relevant variables are defined up front; thus, the state space is generally fixed. Examples of heuristics include breaking a complex problem into sub-problems, hill climbing (reiteratively modifying the current state to look more like the goal state), and working backward from the goal state to the initial state. Heuristic search may include the *restructuring* of mental representations. This restructuring may be accomplished, for example, by (1) re-encoding the problem such that new elements are perceived to be relevant, or (2) relaxing goal constraints (Weisberg, 1995).

The idea that creativity could be construed as a heuristically guided search gave hope to those who sought a scientific understanding of creativity, since search is formally tractable. However, in many creative tasks, and particularly artistic forms of creativity, the goal state is unspecified, and some elements of the eventual solution may not be present when the problem presents itself. It has been suggested that creativity involves heuristics that guide the search for a new state space itself, not just a possibility within a given state space (Boden, 1990; Kaplan & Simon, 1990, Ohlsson, 1992). However, search based approaches to creativity start with pre-existing state spaces and do not address how a new state space comes into existence. Furthermore, like Wallas' four-stage theory, the heuristic search approach to creativity neither addresses the relationship between creativity and hardship nor the therapeutic impact of creativity.

**1.5 Dual Process Theories**

Several proposals for two forms of cognitive processing come from largely disconnected literatures in cognitive and social psychology (Evans, 2008; Sowden, Pringle, & Gabora, 2014). There is interaction between implicit and explicit processes, including synergy effects. Dual processing theories are diverse; while some are concerned with parallel competing processes involving explicit and implicit this repeats from above knowledge systems, others are concerned with the contextualizing and shaping of deliberative reasoning and decision-making by preconscious processes (source?). Nevertheless, these proposals both concern the basic need to distinguish between cognitive processes that are fast, automatic, and unconscious and those that are slow, deliberative, and conscious.

Dating back to Freud's (1949) distinction between primary process and secondary process thinking, most creativity researchers espouse some variant of a dual process theory (e.g., Barron, 1963; Eysenck, 1995; Feist, 1999; Finke, Ward, & Smith, 1992; Fodor, 1995; Gabora, 2003; Martindale, 1995; Richards et al., 1988; Russ, 1993; Simonton, 1999). Psychological theories of creativity typically involve (1) a divergent stage that predominates during idea generation, and (2) a convergent stage that predominates during the refinement, implementation, and testing of an idea (for a review see Runco, 2010). Divergent thought is characterized as intuitive and reflective; also, it involves the generation of multiple discrete, often unconventional possibilities. Divergent thinking ability is sometimes measured in terms of fluency: the number of ideas generated. Convergent thought, characterized as critical and evaluative, involves the selection or tweaking of the most promising possibilities. Neural models of the mechanisms underlying these two modes of thought have been proposed (Gabora, 2002, 2010, 2018; Gabora & Ranjan, 2013).

One well-known dual process theory of creativity is the Geneplore model (Finke et al.,1992). This theory posits that the creative process consists of two stages: generate and explore. (Indeed the name 'Geneplore' is a condensation of "generate" and "explore.") The generation stage involves coming up with crudely formed ideas referred to as pre-inventive structures that contain the kernel of an idea as opposed to an idea in its entirety. The exploration stage involves fleshing out these pre-inventive structures through elaboration and testing.

Use of the term *exploration* to refer to the second phase of the creative process can be misleading. *Explore* is often used to refer to surveying the space of possibilities as generally occurs during the first phase of the creative process, as opposed to refining a single possibility as generally occurs during the second phase. However, the notion of a pre-inventive structure does capture the intuition that early on in the creative process one is working with cognitive structures that are different in kind from those being worked with later in the creative process. The Geneplore model does not attempt to formalize how a pre-inventive structure differs from a full-fledged ideaor what differentiates a promising pre-inventive structure from a mundane one.

Another theory of creativity that could be considered a dual process theory emphasizes ideation-evaluation cycles (Basadur, 1995). Creative thinking is said to involve three major stages—problem finding, problem solving, and solution implementation.Each of these involves alternating cycles ideation and evaluation to varying degrees, depending on the domain. Domains that emphasize problem finding have a higher ratio of ideation to evaluation, whereas domains that emphasize implementation show the opposite.

A dual process theory of analogy is structure mapping (Gentner, 1983). In brief, analogy generation occurs in two steps: first, searching memory in a "structurally blind" manner (Gentner, 2010, p. 753) for an appropriate source and aligning it with the target. Second, mapping the correct one-to-one correspondences between the source and the target.

Yet another well-known dual process theory of creativity is the Darwinian theory of creativity (Campbell 1960; Simonton, 2011). Like biological species, creative ideas exhibit the kind of cumulative complexity and adaptation over time as an evolutionary process, not just when they are expressed to others but in the mind of the idea's creator (Gabora, 1996; Terrell, Hunt, & Gosden, 1997; Thagard, 1980; Tomasello 1996). Thus, it has been proposed that in creativity, as in natural selection, there is a phase conducive to generating variety and another conducive to pruning out inferior variants. According to the Darwinian theory, we generate new ideas through essentially a trial-and-error process involving *blind* generation of ideational *variants* followed by *selective retention* of the fittest variants for development into a finished product. Thus, the Darwinian theory is sometimes referred to as Blind Variation Selective Retention (BVSR). The variants are said to be 'blind' in the sense that the creator has no subjective certainty about whether they are a step in the direction of the final creative product.

In addition to serious theoretical flaws with BVSR (e.g., Gabora, 2007), although the relationship among creativity, hardship, and well-being is at times mentioned in BVSR and other dual process accounts, it does not play a central role in these theories. If we were to find out suddenly that we were wrong about the research relating creativity to hardship and the therapeutic benefits of creativity, these theories would not require substantial revision as a result.

**1.6 Honing Theory (HT)**
This relationship among creativity, hardship, and well-being plays a fundamental role in another theory of creativity known as the honing theory (Gabora, 2017). While the central aim of the above-mentioned theories of creativity is to account for the existence of creative products—i.e., products that are new and useful to society—the central aim of the honing theory of creativity is to account for the cumulative nature of cultural evolution. This is explained not as creative outputs but as the minds that generate them. Thus, HT focuses not just on restructuring as it pertains to the conception of the task, but also as it pertains to the global structure of the mind, what we call the worldview.

A *worldview* is a mind experienced subjectively, from the inside. It is a way of *seeing* and *being in* the world that emerges as a result of the structure of one's web of understandings, beliefs, and attitudes. A worldview reveals itself through behavioural regularities in how it is expressed and responds to situations (Gabora, 2017). The creative process reflects the natural tendency of a worldview to self-organize to achieve a state of dynamical equilibrium through interactions amongst its components, whether they be ideas, attitudes, or bits of knowledge. Most people are familiar with the experience of "catching themselves" in internal dialogue. Internal dialogue is evidence of the self-organization in action of one's worldview.

An evolutionary theory, HT developed from the premise that creativity is the novelty-generating component in cultural evolution. This is an evolutionary process, that is, a process of cumulative, open-ended, adaptive change over time. However, it is not a Darwinian or *selectionist* evolutionary theory (Gabora, 2011). Although *selection* as the term is used in the layperson sense may play a role (i.e., people may be selective about which aspects of their worldviews they express or which paintings they show at a gallery), the process does not involve selection in its technical sense. (This involves change over generations due to the effect of differential selection on the distribution of heritable variation across a population.) As in any kind of evolutionary process, novelty generation must be balanced by novelty preservation. In biological evolution, novelty is generated through genetic mutation and recombination, and the novelty is preserved through the survival and reproduction of "fit" variants. In cultural evolution, novelty's generation is through creativity, and novelty's preservation through imitation and other forms of social learning.

HT posits that the creative process begins with being alert to *psychological entropy*, arenas of one's worldview that, on the spectrum from orderly to chaotic, are relatively chaotic and in need of creative restructuring (Gabora, 2017; Hirsh, Mar, & Peterson, 2012). The process can be "jogged along" by stimuli that capture attention or pique interest; creativity often involves a *seed incident* that gets the creative juices flowing (Doyle, 1998).

Honing an idea involves looking at it from the different angles proffered by one's particular worldview: "putting one's own spin on it," making sense of it in one's own terms, and expressing it outwardly (Gabora, 2017). HT posits that creativity involves viewing the task from a new context, which may restructure the internal conception of it. This restructuring may be amenable to external expression. Thus, honing enables the creator's understanding of the problem or task to shift, and in so doing a form may be found that fits better with the worldview as a whole. In this way, not only does the task get completed (or worked on and put aside) but also the worldview transforms, becoming more robust as it evolves.

The transformative impact of immersion in the creative process extends far beyond the "problem domain." It can bring about sweeping changes to that second (psychological) level of complex, adaptive structure that alters one's self-concept and view of the world. Creative acts and products render such cognitive transformation culturally transmissible. This is why HT posits that what evolves through culture are not creative contributions but worldviews. Cultural contributions give hints about the worldviews that generate them. When faced with a creatively demanding task, not only does one's worldview offers perspectives that alters the conception of a task, but, likewise, immersion in the task subtly or profoundly alters one's worldview. The above-mentioned finding that childhood adversity is a developmental antecedent of creativity is consistent with viewing creativity as the honing and expressing of a unique worldview, since adversity and isolation generate the need and mental space to figure things out for oneself. It is through the creative honing of networks of understandings that worldviews self-organize, and it through the communal expression of honed ideas that culture evolves.

Midway through the creative process, one may have made associations between the current task and previous experiences, but not disambiguated which aspects of those previous experiences are relevant to the current task. At this point, the idea may feel "half-baked." It can be said to be in a *potentiality state* because how it will actualize depends on the particular perspectives from which it is considered. These perspectives may be internally generated (imagining what would happen if …) or externally generated (e.g., building a prototype and trying it out). Thus, the recursive process described in which an external change suggests a new context from which to think about the creative task, and so forth recursively until the task is complete is sometimes referred to as *context-driven actualization of potential*. Each time the idea is looked at from a new context it undergoes a change-of-state such that some of its potential becomes more readily actualized. When the task is complete, the conception of it is said to be in an *eigenstate*, because one's worldview is no longer spontaneously generating new contexts from which to consider it. The creator may express this state as a product, which can cause someone else's worldview to be in a potentiality state. This is when it is the other's turn to adapt it to their own needs or tastes. Through this process culture evolves in new directions.

A worldview not only self-organizes in response to perturbations but it is imperfectly reconstituted and passed down through culture. This is because it is not just self-organizing but *self-regenerating*: people share experiences, ideas, and attitudes, thereby influencing the process by which others' worldviews form and transform. Children expose elements of what was originally an adult's worldview to different experiences and bodily constraints, and thereby forge unique internal models of the relationship between self and world. Thus, worldviews evolve by interleaving (1) internal interactions amongst their parts, and (2) external interactions with others. Through these social interactions, novelty accumulates and culture evolves. Elements of culture create niches for one another. One creative ideas begets another and modifications build on each other, a phenomenon sometimes referred to as the *ratchet effect*.

**1.7 Theory and Research on Mindfulness**
At the beginning of the chapter, we defined mindfulness as the awareness of what is happening presently, by paying attention to our experience from a novel perspective, without judgement. Like creativity, the practice of mindfulness appears to reduce dissonance, enhance feelings of connection, and facilitate the movement of repressed emotion. This correlates with indicators of well-being (the reduction of stress, anxiety, and depression) and it has also been related to satisfaction of life, vitality, a sense of flourishing, and self-actualization (Beitel, Ferrer. & Cecero, 2004; Brown & Codon, 2016; Brown & Ryan, 2003; Cardaciotto, Herbert, Forman, Moitra, & Farrow, 2008; Carlson & Brown, 2005; Feldman, Hayes, Kumar, Greeson, & Laurenceau, 2007; Lawlor, Schonert-Reichl, Gadermann, & Zumbo, 2014; Walach, Buchheld, Buttenmuller, Kleinknecht, & Schmidt, 2006).

If creativity can reduce the dissonance in a worldview, could it then lead to mindfulness, which reduces disharmony in well-being?  If a person knows nothing about what mindfulness is, can he or she in the heart of the creativity become more (unknowingly or knowingly) mindful? If so, does the individual afterward feel less gripped by internally aroused or externally provoked stress, and therefore more connected to others? If this is the case, does this person feel more revitalized, open to possibility (an increase in potentiality), and able to influence social networks towards a creative shift? Perhaps a feedback loop of creativity > mindfulness > creativity might transformatively impact not only an individual's creative process but also her or his psychological structure to adaptably alter self-concept and thus move towards self-actualization.

The impact of mindfulness on creativity is being studied (Baas, Nevicka, Ten Velden, 2014; Greenberg, Reiner & Meiran, 2012; Ostafin & Kassman, 2012; Ren, Luo, Wei, Ying, Ding et al., 2011). Such research connects mindfulness practice, such as focused attention meditation and open monitoring meditation, to divergent and convergent creative thinking. Research also finds a positive correlation with mindfulness meditation on positive insight problem solving, reduction in habitual responses, an increase in cognitive flexibility, fluency and originality (Baas et al., 2014; Capurso, Fabbro, Crescentini, 2014; Chambers, Gullone, & Allen, 2009; Colzato,

Ozturk, & Hommel, 2012; Greenberg, et al., 2012), as well as attentional focus (Davidson & Lutz, 2008; Valentine & Sweet, 1999) and the inhibition of automatic responding (Schmertz, Anderson, & Robins, 2009). With these considerations, mindfulness can be said to lead to creativity.

Our question, however, is can creativity lead to mindfulness? For our purposes here, creativity that is enhanced or engaged in due to a pre or concurrent state of mindfulness we will call *mindful creativity*, which is in common use (Haller, 2015). However, mindfulness that is enhanced or engaged in due to a pre or concurrent process of creativity we will refer to as *creative mindfulness*, which redirects how this term is often used to focus on creativity *leading* to mindfulness. In brief, a person who is being more mindful might become more creative, and someone being more creative might become more mindful.

Some work that could help answer the question of the extent to which creativity can cultivate mindfulness is the research on *flow*. Csikszentmihalyi (2014) defines flow as a "holistic sensation present when we act with total involvement," (p. 136) that is experienced in creativity, but not always or is limited to it. He distinguishes flow from mindfulness in the sense that in flow one is aware of one's actions, but not aware of being aware. For example, one may be aware of painting an artistic scene on a canvas, in part due a narrowing of attention on the canvas, but not consciously aware that one is in a room with a cat in the periphery, focused directly on painting a painting. However, he argues that flow fuels the motivational drivers that lead to creativity (Csikszentmihalyi (2015).

Mindfulness has been described as "the awareness that emerges through paying attention on purpose, in the present moment, and nonjudgmentally to the unfolding of experience moment by moment" (Kabat-Zinn, 2003, p. 145). It also includes "simultaneously drawing novel distinctions in the present moment" (Haller, et al., 2016, p. 894). We posit here that mindfulness has three components (Shapiro & Carlson, 2009; Young, 2016):

1) *Awareness* of what is happening presently, from a subjective (self-referential) and concurrently objective (non-self-referential) point of view (Vázquez Campos & Gutiérrez, 2015).
2) *Attention*, through the novel distinctions in observation compared to the habitual. One might ask "what is new in what I'm observing?" This includes focused attention on a particular object (both externally and internally) and an open attention, which defocuses from a particular object or, rather, is a distributed attentional focus where one is attentive to experience and the interpretation of it
3) *Equanimity*, including the non-judgmental acceptance of what is being experienced and attended to.

Mindfulness can result in observation, non-evaluative description,, acceptance (Baas et al., 2014), as well as a "greater sensitivity to one's environment, more openness to new information, the creation of new categories for structuring perception, and enhanced awareness of multiple perspectives in problem solving" (Langer & Moldoveanu, 2000).

Mindfulness can be contextualized within four modes of perspective. These are *subjective* (thoughts, emotions, perceptions and sensations); *objective* (a time/space material observation of body and externalities); *inter-subjective* (values, relationships, and meaning amongst social connections), and *inter-objective* (systems, networks, and environments), as per integral theory (Forbes, 2016). Integral mindfulness practices create the context of a mindfulness experience within not only an internal (subjective) and external (objective) awareness, but also one that is collective (inter-subjective) and systemic (inter-objective). From the perspective of its systemic mode, mindfulness programs, for example, have been used to engage in social actions around the Occupy movement, and anti-racist, climate change, and social justice initiatives in schools and communities (Magee, 2015; Rowe, 2015a, 2015b; Forbes, 2016). In this way, the systemic mode of mindfulness can lead to social inclusivity, equitable citizenry, communal well-being (eudemonia), and a "shared meaning of the common good" (Forbes, 2016, p. 1267; also, Giroux 2014; Healey 2015a, 2015b).

Although mindfulness has distinct components such as attention, observation, and equanimity, it has been shown that the "ability to carefully observe, notice, or attend to a variety of internal and external phenomena consistently predict[s]enhanced creativity" (Baas et al. 2014, p. 1103). Still, observation has been considered one of the stronger indicators of creativity; it is associated with being open to experience and assisting in adaptive responses to uncertain or complex situations (Siegel & Siegel, 2014). While a mindfulness induced observation leads to improved creativity, it has not been shown conclusively that a creativity induced observation leads to improved mindfulness.

Should a study be conducted in this manner, creativity could be seen to initiate mindfulness, and beyond this possibly a deeper sense of personal awareness and social awareness. A heightened awareness could then lead to a positive shift towards creating transformative environments for individuals and communities to deal with complex problems innovatively. Creative mindfulness, meaning-making, and social capital could result as demonstrated by enhanced trust, community engagement, and meaningful individual and social action (Ponder, 2012). Bridging outward social action with inner transformation in a collective way can enable social and self-actualization. While such cause and effect may not be so direct, such possibilities act as starting points for the conscious honing of the networks of understandings by which worldviews self-organize, which is in part how culture evolves. As Mouchiroud and Bernoussi (2008) conclude: "It may be that only socially creative individuals will be able to act efficiently on these global issues and invent viable social solutions" (p. 379).

**1.8 Duress and Well-being in Theory on Mindfulness and Creativity**
We all want well-being of some kind in our lives. We seek it for ourselves and others in many ways: financial security, support of family members and friends, and in our own personal sense of self. Self-agency, emotional regulation, and higher self-esteem are correlated with mindfulness (Siegel, Siegel, & Parker, 2016), which may help to facilitate our ability to meet basic needs we have for autonomy, competence, and relatedness (Brown & Ryan, 2003).

Research has been done to determine if the Mindfulness-Based Stress Reduction (MBSR) program and the Eight Point Program (EPP), a concentration-based meditation program, decrease duress and enhance well-being for those with medical and psychiatric conditions, as measured by the Mindful Attention and Awareness Scale (MAAS; Brown & Ryan, 2003; Easwaran, 1991). In this study, mindfulness was defined by the non-judgemental awareness of moment-to-moment experiences. Findings suggested significant increases in well-being (Shapiro, Oman, Thoresen, Plante, & Flinders, 2008), with well-being defined by improved effects in perceived stress, rumination, and forgiveness. Another study showed that when well-being was measured for self-esteem, neuroticism, positive or negative affect, self-actualization, autonomy, competence, and physical health, MAAS-measured mindfulness had wide-ranging and inclusive correlations to enhanced well-being (Brown & Ryan, 2003). There is evidence that daily creative activity in everyday environments has a positive impact on emotional experience (Conner, DeYoung & Silva, 2018). This supports the growing perspective that everyday creativity can cultivate positive psychological functioning and thus a sense of flourishing and well-being (Richards, 2010).

Well-being has been shown to be correlated with *social and emotional competence* (SEC; Jennings, 2016), which refers to the ability to manage stress, emotional reactivity, and related cognition and behaviour to optimize effectiveness in daily activity such as the classroom. SEC's five competencies are relationship skills, social awareness, responsible decision-making, self-awareness, and self-regulation. Our external abilities toward social competency are related to our internal abilities toward personal competency. Interpersonal connections, trust, compassion, internal kind-heartedness, and self-awareness all result from mindfulness (Siegel et al, 2016). Although the SEC has been applied to teaching, students may also benefit.

When a contemplative capacity such as mindfulness along with social competence is relatively high, an individual is more likely to be creative (Zajonc, 2014). The type of creativity that instigates or results from social awareness could be called *social creativity* (Watson, 2007), although this term also applies to multiple conditions. *Social mindfulness* has been described as a "benevolent focus on the needs and interests of others. General mindfulness starts with paying attention to the little things available to individual awareness; social mindfulness starts at a similar basic level" (van Doesum, van Lange & van Lange, 2013).

For the sake of this chapter, we propose that social mindfulness is the attentiveness of greater individual awareness as applied to the needs and interests of others. Social mindfulness is founded in both inter-subjective and inter-objective modes of mindfulness. Similarly, social creativity has been related to a higher sense of meaning-making, such as self-actualization (Maslow, 1971; Serotkin, 2010). Social creativity has been used in social awareness and social justice programs, anti-oppression pedagogy, democratic reform, and social advocacy (Beyerbach & Ramalho, 2011; Boal, 2000, 2005; Dewhurst, 2014; Hick & Furlotte, 2009; Lampert, 2013) and it is at the heart of social innovation programs in that such programs offer novel approaches to social problems (Mumford & Moertl, 2003). Both social mindfulness and social creativity increase an individual's awareness of the needs of others, connections within groups and communities, which deepen a sense of meaning and increase personal and social well-being while augmenting social change.

**1.9 Educational Applications of Creativity and Mindfulness Research**
The pace of cultural change is accelerating more quickly than ever. In biological systems, environmental change often induces a sudden increase in the mutation rate. This makes adaptive sense, although most mutations are detrimental while others are beneficial and enhance their survival in changing environments. Similarly, in times of accelerated cultural change it is adaptive to increase the rate of creativity and generate innovative solutions to any unforeseen problems. This is particularly important now; in our high-stimulation environment, children spend so much time processing new stimuli that there is less time to go deep with the stimuli they have encountered. They also do not have much time to think about ideas and situations from different perspectives, such that their ideas become more interconnected and their mental models of understanding more integrated. It is this kind of deep processing and the resulting integrated webs of understanding that make the crucial connections leading to important advances and innovations.

Therefore, given the fast pace of change in society today we are led to ask, how can creativity and mindfulness be cultivated in the classroom? Starting with creativity, one way is by focusing less on the reproduction of information, and more on problem solving and critical thinking. Another way of cultivating creativity in this context is by posing questions and challenges, followed by experiencing opportunities for solitude and reflection or group discussion in an effort to foster the honing of new approaches and ideas.

In addition, to incorporate teaching methods that encourage creativity, we suggest that classroom curricula might include a regular Creativity and Mindfulness Module that) includes activities and approaches inspired by the creativity and mindfulness literatures. Creativity-enhancing activities can be, for example, assignments that transcend traditional disciplinary boundaries. Examples include painting murals that depict historical events, acting out plays about animals that share a biological ecosystem, and writing poetry about black holes. After all, the world does not come carved up into different subject areas. It is only through enculturation that we come to believe these disciplinary boundaries are real—our thinking becomes trapped in and by them.

There are also ways in which mindfulness can be cultivated in the classroom both on a day-to-day basis and as part of a possible Creativity and Mindfulness Module. If formal mindfulness programs are not in place or practical, a *creative mindfulness* approach would offer students the benefits of creative engagement and possibly heightened well-being. This can be accomplished with students through emphasis (general or specific) on being creative in present-moment awareness, by providing the physical and psychological space for full engagement in the creative act (e.g., in art, science, physical education, etc.) via awareness, attention, and equanimity. For example, a teacher could guide students into focusing their attention on the details of what a student is doing by not only paying attention to their task, but also their process and environment (being aware of one's surroundings as well as the study at hand in a balanced way), and not judging themselves or others' experience or outcome. By helping students 'get lost' in their creative experience, to the point of them being engrossed with the very act of creating itself, teachers may find their students become more involved as whole beings, more able to pay attention with greater awareness, and overall more mindful as a result.

Creative mindfulness, when directed to bring the group into the heart of the creative process in any domain, could lead to self-awareness and then to social awareness of the group dynamic. If focused, this could bring about peak experiences such as flow that are rich with a sense of vitality and total involvement. As William James (1923), acclaimed father of psychology, claimed, "the faculty of voluntarily bringing back a wandering attention, over and over again, is the very root of judgment, character, and will. … An education which should improve this faculty would be *the* education *par excellence*'' (p. 424, italics in original). With these considerations, youth can be prepared to thrive with uncertainty, open their minds to the novelty found in observing themselves and their studies with awareness, while nurturing personal and social well-being.

**1.10 Conclusions**
Creativity is much-needed in today's fast-paced, ever-changing world. In the long term, it can be transformative, therapeutic, and even self-actualizing. In the short term, however, creativity can re-open painful memories, causing duress. Yet, there is reason to believe that—paired with the potentially life-affirmating effects of mindfulness towards well-being—it may be possible to delve into painful experiences and tendentious material but nevertheless feel held and accepted by one's higher self throughout the process. With this type of self-regulation, creativity may help one improve a generalized sense of self-awareness and even social mindfulness, as the needs of others and connections to community, meaning-making, and social well-being develop. Such changes could add to a possible collective shift in social-actualization and a potential change in worldview. Drawing upon the literatures in the psychology of creativity and mindfulness, we have loosely sketched out a creativity and mindfulness module that introduces a *creative mindfulness* approach, for prospective application in a classroom setting. This is just an initial outline that requires development and refinement. Nevertheless, we suggest that a move in this direction could play a part in an educational curriculum designed to address unique challenges of the 21st Century.

**Acknowledgements**
This research was supported in part by a grant (62R06523) from the Natural Sciences and Engineering Research Council of Canada.